# Fluid-structure model for disks vibrating at ultra-high frequency in a compressible viscous fluid

H. Neshasteh,[1] M. Ravaro,[1] and I. Favero[1,*]

[1] *Matériaux et Phénomènes Quantiques, Université Paris Cité, CNRS UMR 7162, Paris, France*

[*] Corresponding author: ivan.favero@u-paris.fr

## Abstract

Radial mechanical modes of miniature disk-shaped resonators are promising candidates for probing the ultra-high-frequency rheological properties of liquids. However, the lack of an analytical fluid-structure model hinders the inference of liquid properties from the experimental measurement of such radial vibrations. Here we develop analytical models for the case of a disk vibrating in a compressible viscous liquid. Closed-form expressions for the mechanical quality factor and resonant frequency shift upon immersion are obtained, which we compare to the results of numerical modeling for a few significant cases. At frequencies above 1 GHz, our model points out the significance of compressibility effects.

The direct mechanical probing of liquids at high frequencies still constitutes a scientific and technological challenge. Existing approaches are sparse, with a frequency gap between fast mechanical rheometers operating at tens of MHz (<100MHz)[1,2] and recent pump-probe experiments on nanostructures vibrating between 10 and 20 GHz[3-5]. The development of micro- and nanomechanical systems has recently allowed exploring basic questions in hydrodynamics, by revisiting at small scale the paradigm of a solid body vibrating in a fluid[6–8]. These systems also led to new applications such as high-throughput rheology[9] and in-situ particle detection[10,11]. These progresses were made at frequencies of tens of MHz at most. Upon further miniaturization and device design, it is only recently that ultra-high frequency (UHF) mechanical probes (100MHz-5GHz) could be measured while vibrating in liquids, using the high-sensitivity permitted by optomechanical techniques[12,13]. However, complete models that describe the solid-liquid interaction in this latter regime are still limited. It is the subject of the present paper to improve this current state.

In spite of their miniaturization, the dimensions of UHF mechanical resonators obtained by top down fabrication (100 nanometers to a few microns) remain orders of magnitude larger than the mean free-path in a liquid (about 2nm in water). This justifies treating the liquid as a continuum, which under some approximation for the deviatoric stress leads to a Navier-Stokes form for the hydrodynamics equations[7,14]. Solving these equations analytically is in general a formidable task, and one often has to resort to numerical modeling instead. Exact solutions exist for the simplest geometries, such as a sphere[14], and in the last decades approximate solutions were developed for cases of practical interest such as circular plates[15–17], fluidic channels[18], or rectangular cantilevers[19–22]. These works, however, did consider the liquid as incompressible, with the liquid viscosity being the only source of dissipation. While reasonable for the considered resonators, this approximation may not hold for other geometries of interest where the mechanical wavelength becomes commensurate with the length scale of the flow[23,24].

A compressible fluid being elastic, it requires the concomitant treatment of compressibility and viscous effects and hence falls under the category of visco-elastic problems, even if relaxation effects in the stress are absent in the constitutive stress-strain relaxations. In a seminal paper[25], the problem of a sphere oscillating in a compressible viscous liquid was solved exactly using a Voigt body description for elastic and viscous media. Exact solutions were then derived for an infinite cylinder vibrating in a compressible viscous liquid[26]. Compressibility and viscosity effects were recently observed on micron-sized optomechanical disk resonators vibrating radially in the UHF range[12], in an effort to export optomechanics experiments into liquid environments[12,13,27,28]. A first analytical fluid-structure model was proposed[12] to describe these observations, and confronted to experiments and numerical predictions. Inspired by the sphere problem[25], this model led to closed-form expressions for the mechanical frequency and damping of the resonators in the limit of an incompressible viscous fluid, and was extended by mere empirical expressions in the limit of an inviscid compressible fluid[12]. A gap remained between these two limits, precluding a complete analytical description when both viscosity and compressibility were at play. The aim of the present paper is to fill this gap and provide now an analytical description that is valid in all regimes of the surrounding liquid. Under proper approximations, we additionally obtain here closed-form expressions for the mechanical frequency and damping of the resonators. These expressions enable a consistent extraction of elastic and viscous properties of the liquid from a linear mechanical spectroscopy of the resonator.

For an isotropic compressible viscous liquid, the constitutive equation for the fluidic stress $\sigma$ reads[14]:

$$\sigma = -p\mathbb{I} + 2\mu\left[\frac{dS}{dt} - \frac{1}{3}(\text{div }\mathbf{v})\mathbb{I}\right] + \mu_B(\text{div }\mathbf{v})\mathbb{I} \qquad (1)$$

with p the pressure, $\mu$ the shear viscosity, S the strain tensor, $\mathbf{v}$ the velocity and $\mu_B$ the bulk viscosity. The elastic response results from the compressibility $\beta=\frac{1}{\rho}\frac{\partial\rho}{\partial p}$, with $\rho$ the density, which implies $\frac{\partial p}{\partial t} = -\frac{1}{\beta}\text{div }\mathbf{v}$. Since div $\mathbf{u}$ = Tr(S), and hence div $\mathbf{v} = \frac{\partial \text{Tr}(S)}{\partial t}$, the stress in Eq. 1 is a specific linear function of the strain and of its time-derivative, in the spirit of a Voigt body description. In order to depict our UHF resonator, we adopt a one-dimensional lumped element mechanical model of the resonator. In vacuum, it has a resonant angular frequency $\omega_s^2$=$k_s$/$m_s$, with $k_s$ the (solid) spring constant and $m_s$ the (solid) mass. In presence of a compressible viscous liquid, the equivalent mechanical circuit is represented in Fig. 1 (a), where elastic and viscous effects in the fluid appear as parallel elements, in agreement with a Voigt model. The spring is inversely proportional to $\beta$, and we explicitly separate in the circuit the shear from bulk viscosity, in order to distinguish contributions from shear and longitudinal motion. Once in liquid, the resonator frequency is shifted from $\omega_s$ to $\omega_l$, while the resonance line acquires a finite width, as represented in Fig. 1 (b).

In presence of the fluid, the evolution of the resonator mechanical energy $E_m$ is linked to the mechanical power provided by the fluid to the vibrating disk[29]:

$$\frac{dE_m}{dt} = \int v_j \, \sigma_{ij} n_i d\Sigma \qquad (2)$$

with n the unitary vector normal to the elementary resonator-fluid surface $d\Sigma$. Taking the time derivative of (2) and moving into complex notations in case of harmonic oscillation $x = \text{Re}(\tilde{x} \, e^{j\omega t})$, the equation of motion becomes:

$$(-m_r \omega^2 + j\gamma\omega + \omega_s^2)\tilde{x} = 0 \qquad (3)$$

where the fluid-structure interaction results in an effective harmonic resonator with mass $m = m_r m_s$, damping rate $\gamma$, and a loaded complex frequency $\tilde{\omega}$ that is solution of the polynomial Eq. 3 with:

$$\gamma = \frac{-\text{Re}\left(\int \tilde{v}_j \tilde{\sigma}_{ij} n_i d\Sigma\right)}{2E_s} \qquad (4)$$

and

$$m_r = \left(1 - \frac{\text{Im}\left(\int \tilde{v}_j \tilde{\sigma}_{ij} n_i d\Sigma\right)}{2\omega_s E_s}\right) \qquad (5)$$

with $E_S$ the total mechanical energy of the bare resonator, and where the choice of phase reference is such that $\tilde{v}$ is real. For simplicity, the tilde notation will be dropped in the followings, if not explicitly required.

Eq. 4, 5 and 1 show that the knowledge of the fluid velocity at the resonator interfaces is required to obtain both the frequency shift and added damping upon liquid immersion. In the following we show that this information can be obtained through a perturbative mode-matching approach: first, the resonator vibration profile is calculated in vacuum, in absence of the fluid; second, the velocity of the fluid is calculated by solving hydrodynamic equations and imposing to match the bare resonator's velocity at interfaces. This approach is expected to be satisfactory as long as the vibration profile is not strongly modified by the presence of the fluid, and tractable provided that efficient approximations can be employed at interfaces.

The benchmark system we chose to test this approach is a suspended micron–sized disk resonator, such as those recently operated in liquids through optomechanical techniques[12,27,30,31]. When coming to numerical illustrations below, we will specifically address the case of silicon disks with a radius ranging from 1 to 10 µm, fabricated out of a 220nm thick layer of <100> silicon of density $\rho_s$, Young modulus E and Poisson ratio ν. Among the variety of mechanical modes of such resonators, we focus on the fundamental radial breathing mode (RBM), with has a dominant radial displacement, shown in Fig. 1 (c), and is known to provide strong optomechanical interactions[32]. The disk is considered to be immersed in a compressible Newtonian fluid of density ρ, compressibility β, and shear (bulk) viscosity µ ($\mu_B$). In the isotropic elastic solid assumption taken here, the fundamental RBM profile is invariant upon azimuthal rotation around the disk axis, and the displacement inside the solid can be expressed in cylindrical coordinates as $\mathbf{u}(r,z) = u_r(r,z)\mathbf{r}+u_z(r,z)\mathbf{z}$, with $|u_r| \gg |u_z|$ and **r** and **z** unit vectors. The same holds for the velocity field in the surrounding isotropic fluid $\mathbf{v}(r,z) = v_r(r,z)\mathbf{r}+v_z(r,z)\mathbf{z}$.

For small displacements, the displacement profile **u**(r,z) of the fundamental RBM in vacuum is obtained from elasticity theory for an isotropic, homogenous elastic solid. In the limit of a thin plate disk (h ≪ a) and in absence of stress exerted on the top and bottom surfaces, one can approximate the stress components in the solid $\sigma_{zz}, \sigma_{rz} \approx 0$. The rotation invariance of the fundamental RBM also implies $\sigma_{z\varphi} = 0$ (and $\sigma_{r\varphi} = 0 = S_{r\varphi} = S_{z\varphi}$), eventually leading to a stress in the (r, $\varphi$) plane. The wave equation $\rho_s \frac{\partial^2 \mathbf{u}}{\partial t^2} = (\lambda + 2\mu)\boldsymbol{\nabla}(\boldsymbol{\nabla} \cdot \mathbf{u}) - \mu \boldsymbol{\nabla} \times \boldsymbol{\nabla} \times \mathbf{u}$ can hence been written in the (r, $\varphi$) plane, with $\lambda$ and µ the 2D Lame coefficients (see supplementary material). The RBM being compressional in nature, we look for a curl-free wave $\mathbf{u} = -\boldsymbol{\nabla}\phi$, with φ a scalar potential. At an angular frequency ω, this leads to a 2D wave equation:

$$\nabla^2 \phi + k_s^2 \phi = 0 \; ; \; k_s = \omega_s \sqrt{\frac{(1-\nu^2)\rho_s}{E}} \qquad (6)$$

with $k_s$ the wave vector. The equation is solved by $\phi = A\,J_0(k_s r)$, corresponding to a 2D displacement $\mathbf{u} = -\boldsymbol{\nabla}\phi = A k_s J_1(k_s r)\mathbf{r}$. In 3D, and in the limit of a thin disk, the radial displacement is independent of z and the total displacement is obtained through the stress-plane relation $S_{zz} = -\frac{\nu}{1-\nu}(S_{rr} + S_{\varphi\varphi})$, imposing $u_z=0$ at z=0:

$$\mathbf{u} = Ak_s J_1(k_s r)\,\mathbf{r} - \frac{\nu}{1-\nu} A k_s^2 J_0(k_s r) z\,\mathbf{z} \tag{7}$$

By imposing a stress-free boundary condition $\sigma_{rr} = 0$ on the disk sidewall (r=a), we obtain in this limit a characteristic equation for the wave vector:

$$k_s^2 J_0(k_s a) + (\nu - 1)k_s \frac{J_1(k_s a)}{a} = 0 \tag{8}$$

The mechanical energy stored in the bare vibrating disk, of interest to numerically illustrate Eqs. 4 and 5, is derived from the displacement and frequency extracted from Eqs. 7 and 8:

$$E_s = \frac{1}{2}\rho_s \omega_s^2 \int |\mathbf{u}|^2 dV \tag{9}$$

We move next to the velocity field in the fluid $\mathbf{v}(r,z)$. Neglecting convective terms under the assumption of small amplitude, the field is a solution of the linearized Navier-Stokes equation for a compressible viscous fluid and of a mass conservation relation:

$$j\omega\rho\mathbf{v} = -\nabla p + \mu\nabla^2 \mathbf{v} + \left(\frac{1}{3}\mu + \mu_B\right)\nabla(\nabla \cdot \mathbf{v}) \tag{10}$$

$$j\omega\beta p + \nabla \cdot \mathbf{v} = 0 \tag{11}$$

where the compressibility is linked to the speed of sound in the liquid c through the relation $\beta=(\rho c^2)^{-1}$. We employ an Helmholtz decomposition to express the fluid velocity as the sum of a divergence-free and a curl-free term $\mathbf{v}(r,z)= \nabla\times\mathbf{\Psi}+\nabla\phi$, leading to two independent equations for the scalar and vector potential:

$$\nabla^2\phi + k^2\phi = 0;\quad k = \omega\sqrt{\frac{\beta\rho}{1+j\omega\beta\left(\frac{4}{3}\mu+\mu_B\right)}} \tag{12}$$

$$\nabla^2\mathbf{\Psi} + \sigma^2\mathbf{\Psi} = \mathbf{0};\quad \sigma^2 = -\frac{j\omega\rho}{\mu} \tag{13}$$

This decomposition enables singling-out the contribution of the fluid viscosity and compressibility in fluid-structure interactions. In an inviscid ($\mu=\mu_B=0$) compressible fluid, Eq. 13 implies $\mathbf{\Psi=0}$ and the interactions are restricted to the radiation of curl-free waves. In an incompressible ($\beta=0$) viscous fluid in contrast, divergence-free waves are centrally involved, but curl-free waves are also present and abide $\nabla^2\phi =0$ (hence do not propagate). In what follows Eqs.

12 and 13 are solved concomitantly by means of the perturbative mode-matching approach discussed above. The results are then compared with numerical modeling using the finite element method (FEM), and the general case of a compressible viscous fluid is finally addressed using the superposition principle.

For an inviscid compressible liquid, the velocity field restricts to the curl-free component **v**(r,z)=∇$\phi$. The associated compression waves are essentially driven by the radial displacement of the disk sidewalls, as a consequence of the smaller z-displacement of the disk interfaces predicted by Eq. 7. Fig. 2(a) shows the fluid radial velocity profile obtained by FEM in such limit, in the case of water taken as inviscid and for a silicon disk of 1μm radius and 220 nm thickness. In order to solve Eq. 12 in our model, $\phi$ is expressed separately in both domains I (r>a) and II (r<a) appearing in Fig. 2(a), using an integral spatial expansion on cosine functions in the z-direction, and a separation of variables r and z. Inserting such form in Eq. 12 leads to Eqs. 14 and 15. Zero-order Hankel functions of the second kind appear in the solution in domain I, where compression waves propagate outwards, while zero-order Bessel functions are involved in domain II (Eq. 15) where the velocity field is modulated by a radial interference pattern:

$$\phi_I = \int_0^\infty A(k_z) H_0^{(2)}\left(\sqrt{k^2 - k_z^2}\, r\right) \cos(k_z z) dk_z \tag{14}$$

$$\phi_{II} = \int_0^\infty B(k_z) J_0\left(\sqrt{k^2 - k_z^2}\, r\right) \cos\left(k_z \left(z - \frac{h}{2}\right)\right) dk_z \tag{15}$$

$A(k_z)$ and $B(k_z)$ are calculated by imposing specific boundary conditions: first the continuity of $v_r(r,z)$ and $v_z(r,z)$ between domain I and II both at the solid-liquid interface ($r = a, |z|<h/2$, Eq.16) and within the liquid ($r = a, |z|>h/2$, Eq.17); and second the continuity of $v_z(r,z)$ at the top and bottom surfaces of the disk $v_z(r,z)=0$ (Eq. 18), the last approximation being justified by the smallness of vertical against radial displacement of the disk (see Eq. 7):

$$\left(\frac{\partial \phi_I}{\partial r}\right)_{r=a, |z|<h/2} = j\omega u_r \tag{16}$$

$$\left(\frac{\partial \phi_I}{\partial r}\right), \left(\frac{\partial \phi_I}{\partial z}\right)_{r=a, |z|>h/2} = \left(\frac{\partial \phi_{II}}{\partial r}\right), \left(\frac{\partial \phi_{II}}{\partial z}\right)_{r=a, |z|>h/2} \tag{17}$$

$$\left(\frac{\partial \phi_{II}}{\partial z}\right)_{r<a, |z|=h/2} = 0 \tag{18}$$

Substituting Eqs. 14 and 15 into Eqs. 16-18 results in a system of equations that uniquely determine A($k_z$) and B($k_z$) for all $k_z$, and lead to an analytical expression for **v**(r,z). In practice, we select an appropriate discretization and truncation of the $k_z$ vector spectrum to ensure efficient convergence of calculations. The velocity field obtained this way agrees well with the results obtained with FEM. It notably retrieves the acoustic interference pattern resulting for the RBM source of vibration, which can is well visible in Fig. 2b (FEM results in dotted lines) and resembles that of a radially vibrating ring of radius a. With **v**(r,z) in hand, we then obtain the in-liquid quality factor Q≃$\omega_l$/2Im($\tilde{\omega}$) and loaded frequency $\omega_l = Re(\tilde{\omega})$ of the resonator by expressing the pressure field p(r,z) using Eq. 11, which fully determines the fluidic stress (Eq. 1) required in Eqs. 3-5.

The results of this analytical modeling are illustrated in Fig. 2c and 2d. Both quality factor Q and normalized frequency shift ($\omega_s$-$\omega_l$)/$\omega_s$ are calculated for disk resonators of increasing radius, immersed in various liquids that include water and four distinct perfluorinated liquids (Table 1). All liquids are taken here as inviscid in order to test our model in this limit case. The analytical results show very good agreement with FEM modeling. In Fig. 2c the RBM acoustic radiation Q monotonously increases with the radius, for all liquids. In contrast in Fig. 2d, the frequency shift is peaked, with a peak position that depends on the liquid acoustic properties. In water the peak position sits below 1 μm (frequency of 2 GHz), while it is located at 3 μm for the perfluorinated liquids (frequency slightly below 1GHz).

**Table1. Mechanical properties the investigated liquids.**

|       | ρ (kg/m³) | μ (mPa·s) | c (m/s) |
|-------|-----------|-----------|---------|
| Water | 1000      | 1         | 1500    |
| HT110 | 1730      | 1.57      | 610     |
| HT170 | 1793      | 4.05      | 668     |
| HT230 | 1842      | 11.23     | 702     |
| HT270 | 1868      | 38.6      | 683     |

While the presented semi-analytical approach depicts quantitatively very well the fluid-structure interaction, it does not provide direct closed-form expressions for the quantities of interest such as Q and the frequency shift. In a compressible inviscid liquid, a closed-form expression for Q can in contrast be obtained by approximating the disk by a flat circular ring constituted of an homogeneous distribution of sources of spherical waves (see Supporting Information and figure S1). With such approach we obtain a satisfactory representation of the waves emitted in the far-field visible in Fig. 2b, while we reach an expression for the quality factor of the resonator in the liquid (see Supporting Information):

$$Q = \frac{2c\rho_s}{\rho h \omega_s} \left( \frac{\left(1 - \frac{J_0(k_s a) J_2(k_s a)}{J_1^2(k_s a)}\right)}{\int_0^\pi J_0^2(ka\sin(\theta))\sin(\theta)d\theta} \right) \qquad (19)$$

with $k = \omega_s/c$ the wave vector of the acoustic wave in the fluid. While approximated, this closed-form expression leads to a satisfactory agreement in the compressible inviscid regime, with errors in Q of 30 % in the worst cases investigated (figure S2). In the Supporting Information, we also show how to obtain a closed-form expression for the frequency shift using a similar approach, and we test the validity of this expression by confronting its predictions to the above first analytical approach and to FEM results (figure S2).

For an incompressible viscous fluid, the velocity field includes both terms of the Helmholtz decomposition. The scalar curl-free potential $\phi$ is solution of $\nabla^2 \phi = 0$ (Eq. 12 with k=0), which is solved using the same modal expansion procedure as in the inviscid liquid case. For k=0 the argument of the Bessel and Hankel functions of Eqs. 14 and 15 becomes imaginary with a negative imaginary part, such that there is no propagating wave generated in the fluid in the radial direction, and the radial velocity associated to $\phi$ decreases exponentially for r > a. Note that since $\beta = 0$, $1 + j\omega\beta \left(\frac{4}{3}\mu + \mu_B\right) = 1$ and there is no effect of the bulk viscosity $\mu_B$. At the same time the vector potential $\boldsymbol{\Psi}$ is solution of the heat equation (Eq. 13). Since we are interested in fluidic interactions with the fundamental RBM mode, we look for solutions that have no azimuthal dependence. The velocity field has neither azimuthal component nor azimuthal dependence, and we can hence opt for a vector potential that is strictly oriented along the azimuthal direction, The radial velocity profile shown in Fig. 3a is calculated by FEM for a silicon disk of radius 1µm and thickness 220 nm in viscous water, taken as incompressible. It shows that the mechanical energy emitted by the disk upon vibrating is completely absorbed in the direct vicinity of the resonator, due to strong viscous dissipation. As a consequence the potentials can be described in domains I and II by a single mode. For the azimuthal component of the vector potential, this leads to the following expressions (see Supporting Information):

$$\psi_I = A_I H_1^{(2)}(k_{r_I} r) \sin(k_{z_I} z) \tag{20}$$

$$\psi_{II} = A_{II} J_1(k_{r_{II}} r) e^{-k_{z_{II}}(|z|-\frac{h}{2})} \tag{21}$$

As a consequence of the viscous dissipation and exponential decrease of the radial velocity associated to $\phi$, the total radial velocity at the interface between the distinct fluid domains is negligible for $|z| > h/2$ (Fig. 3a). Hence the only boundary conditions to be fulfilled are at the solid-liquid interfaces. In contrast to the inviscid case, the presence of viscosity invites here to consider (tangential) no-slip conditions at these interfaces, such that four continuity relations need to be considered instead of two in the inviscid case. The continuity of normal velocity on the disk sidewall and top and bottom surfaces remain granted by the curl free component $\phi$ (Eqs. 16 and 18) with a vanishing contribution of the $\Psi$ component, hence fixing a unique solution for $\phi$, while the continuity of the tangential velocity at interfaces involves the two components $\phi$ and $\Psi$, fixing a unique solution for $\Psi$:

$$\left(-\frac{\partial \psi_{II}}{\partial z}\right)_{r<a, |z|=h/2} + \left(\frac{\partial \phi_{II}}{\partial r}\right)_{r<a, |z|=h/2} = (j\omega_s u_r)_{r<a, |z|=h/2} \tag{22}$$

$$\left(\frac{1}{r}\frac{\partial (r\psi_I)}{\partial r}\right)_{r=a, |z|<h/2} + \left(\frac{\partial \phi_I}{\partial z}\right)_{r=a, |z|<h/2} = (j\omega_s u_z)_{r=a, |z|<h/2} = 0 \tag{23}$$

In (23) we again approximate the vertical (z) velocity on the disk surfaces to be zero, because it is much smaller than the radial component (Eq. 7). The total velocity solution found this way has its radial and vertical behavior plotted in Fig. 3b and c for the specific silicon disk resonators considered above. A very good agreement with the results of FEM modeling of the velocity is observed.

In order to express the frequency shift and quality factor of the dressed mechanical resonance, we need to calculate the expressions appearing in Eqs. (4) and (5). The results for Q are shown in Fig. 3d, for the same set of liquids and the same disk radiuses as in Fig. 2c. The normalized frequency shift shown in Fig.3e results from the contribution of both the $\nabla \times \Psi$ component, which is dominating at low frequency (large radius) and monotonously decreasing, and the $\nabla\phi$ component, which instead emerges at high frequency (small radius). There again, the agreement of our analytical approach of the viscous incompressible regime with FEM is very satisfactory.

In the Supporting Information, we go one step of simplification further in order to obtain a closed-form expression, taking notably $k_{r_I} \approx \sigma$. We simplify Eq. (22) by neglecting, on the disk top and bottom surfaces, the radial component of the curl-free component of the velocity (see Fig. S5 for justification), and we finally obtain for an incompressible viscous liquid:

$$\frac{\left(\int \tilde{v}_j \tilde{\sigma}_{ij} n_i d\Sigma\right)}{2\omega_s E_s} = \frac{-2j\mu\sigma}{\omega_s \rho_s h} - \frac{2\mu \frac{\sigma}{\omega_s} \frac{\left(H_2^{(2)}(\sigma a) - H_0^{(2)}(\sigma a)\right)}{H_1^{(2)}(\sigma a)} + \frac{j\rho h}{\pi} ln\left(\frac{8a}{r_0}\right)}{a\rho_s \left(1 - \frac{J_0(k_s a) J_2(k_s a)}{J_1^2(k_s a)}\right)} \quad (24)$$

, reminding that $\sigma^2 = -j\omega_s/\rho\mu$. The validity of this closed-form expression turns out to be very good when taking $r_0 = h/4$. It is discussed at length in the Supporting Information.

Let us finally discuss the case of a compressible viscous liquid. Note that since $j\omega\beta\left(\frac{4}{3}\mu + \mu_B\right) \ll 1$ for typical values of $\mu_B$, we can again neglect the effects of bulk viscosity and assume for example $\mu_B = 0$ in order to derive convenient expressions. An extension to the case of a finite mu_B is proposed in the supplementary material. In order to calculate the velocity by superposing the orthogonal components $\nabla\times\Psi$ and $\nabla\phi$, solutions of Eq. 12-13, the latter need again to respect consistent boundary conditions. In the approximation of a thin disk (h<<a), we neglect the normal components of $\nabla\times\Psi$ on the sidewall, top and bottom surfaces, for the reasons already discussed in the viscous case. The boundary conditions on the normal velocity are then solely imposed by $\nabla\phi$, while those on tangential velocity involve both $\nabla\phi$ and $\nabla\times\Psi$ (Eqs. 22 and 23). This fixes a unique solution, and we can superpose $\nabla\times\Psi$ and $\nabla\phi$ in order to calculate the complex power (Eq.2). The resulting quality factor and normalized frequency shift are shown in Fig. 4 for the same set of liquids and disk radiuses as above. By comparing these results to Fig. 3d-e, we see that for a large disk radius (low frequency), and in particular for the most viscous fluids, the approximation of an incompressible viscous liquid is good enough to understand dissipation. In contrast, for small radius (high frequency), the effects of radiation losses become significant and the quality factor deviates from the incompressible viscous liquid approximation. For the three less viscous liquids of our set (water, HT110 and HT170), there is a minimum for the quality factor as function of radius (for water, the minimum is at smaller radius, out of the shown interval). This minimum corresponds to an acoustic matching condition between the vibrating source and the surrounding liquid. As far as the frequency shift is concerned, the trend is a monotonous decrease of the shift for decreasing radius (increasing frequency). Moreover the difference between the different liquids progressively reduces at smaller radius (higher frequency), pointing towards a reduced impact of viscosity on the shift, while the effects of fluid compressibility start to dominate. This all demonstrates the importance of taking into account the full visco-compressible nature of fluid-structure interactions in order to understand the behavior of UHF mechanical disks vibrating in liquids. In the Supporting Information, we also show how to obtain a useful closed-form expression in such compressible viscous case

$$\frac{\left(\int \tilde{v}_j \tilde{\sigma}_{ij} n_i d\Sigma\right)}{2\omega_s E_s} = \frac{-2j\mu\sigma}{\omega_s \rho_s h} - \frac{2\mu \frac{\sigma}{\omega_s} \frac{\left(H_2^{(2)}(\sigma a) - H_0^{(2)}(\sigma a)\right)}{H_1^{(2)}(\sigma a)} + \rho h \left(\frac{j}{\pi} ln\left(\frac{8a}{r_0}\right) + \frac{1}{2}\int_0^{2ka}(J_0(x) - jH_0(x))dx\right)}{a\rho_s \left(1 - \frac{J_0(k_s a)J_2(k_s a)}{J_1^2(k_s a)}\right)}$$

(25)

and systematically compare the outcome of this closed-form approximation with FEM results, taking $r_0 = h/4$. Corresponding simple expressions for $m_r$ and gamma are proposed in the supplementary material.

In this paper we developed a semi-analytical model to study the mechanical response of a thin disk resonator radially vibrating in a liquid, with frequency in the UHF range. Employing several distinct liquids and disk dimensions, the validity of the model was first tested in the two limit cases of an inviscid compressible fluid (acoustic regime) and in the case of an incompressible viscous fluid (viscous regime). We then discussed the general case of a compressible viscous fluid, and obtained here as well a solid agreement between the semi-analytical model and the results of numerical simulations. In a second stage of approximation, we derived closed-form expressions for both the quality factor and frequency shift of the mechanical resonance of the disk. We confronted these approximations with the two former approaches (semi-analytical and FEM), obtaining satisfactory agreement again. These closed-form expressions bring about the possibility to infer mechanical properties of the liquid (density, compressibility and viscosity) from the linear mechanical spectroscopy of the vibrating disk. The path opened by this theoretical development is that of a controlled rheology of liquids with miniature disk resonators, at a spatial scale (micro-nano) and in a frequency range (UHF) that have remained hard to explore so far.

## SUPPLEMENTARY MATERIAL

In the Supplementary Material, the reader will find. 1) A approximate analytical treatment in 3D of the elastic problem of Radial Breathing Modes of an elastic disk. 2-4) Approximate analytical models for fluid-structure interactions in the inviscid compressible case, leading to closed form expressions for the mechanical frequency shift and quality factor of the disk. 5) An approximate analytical model for fluid-structure interactions in the viscous incompressible case. 6) An approximate analytical model for fluid-structure interactions in the viscous compressible case. 7) A discussion of the case $\mu_B \neq 0$. 8) Technical details regarding FEM simulations.

## ACKNOWLEDGMENTS

This work was supported by the European Research Council via the NOMLI project (n°770933) and by the French Agence Nationale de la Recherche via the HERMES project (n°CE42-0031).

# FIGURES

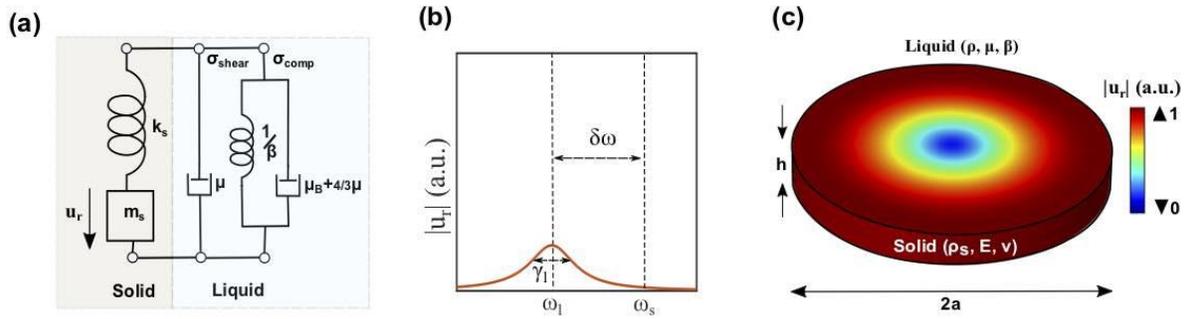

**FIG. 1.** Fluid-structure interaction model for a vibrating disk. (a) One dimensional lump model of a solid mechanical resonator (mass $m_s$ on a spring $k_s$) immersed in a compressible viscous liquid of shear (bulk) viscosity $\mu$ ($\mu_B$) and compressibility $\beta$. (b) Spectral signature of the change in center frequency ($\omega$) and damping ($\gamma$) upon immersion of the resonator into the liquid. (c) Distribution of radial displacement $u_r$ for the first RBM of a disk resonator.

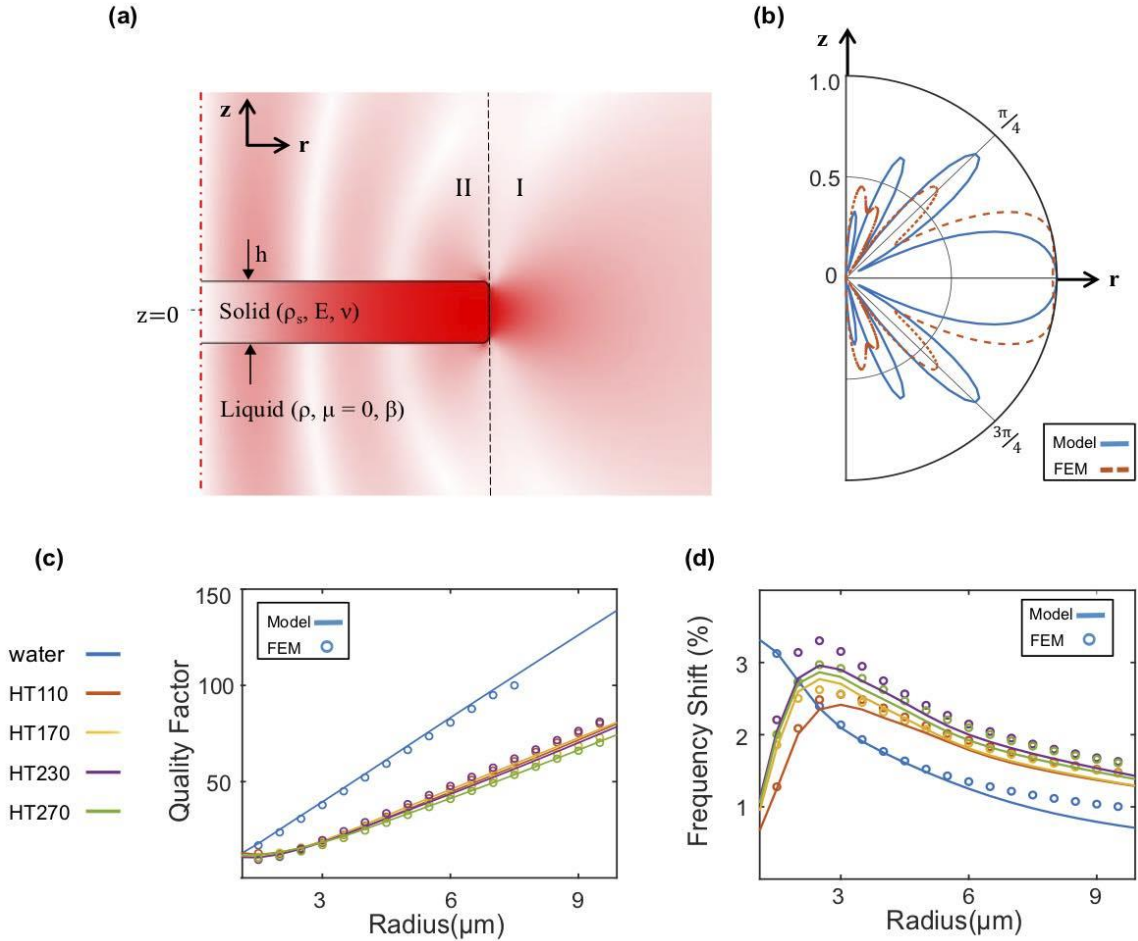

**FIG. 2.** Disk fluid-structure interaction in the inviscid fluid regime. (a) Absolute value of the radial velocity. The considered situation is a silicon disk of radius $a = 1\mu m$ and thickness $h = 220nm$, oscillating on its fundamental RBM and immersed in water taken as inviscid. (b) Analytical (solid line) and numerical FEM calculations (dotted line) absolute value of the radial velocity in far field, for the same disk and liquid as in (a). (c),(d) Analytical (solid line) and numerical FEM (open symbols) calculations of the quality factor and normalized frequency shift of the fundamental RBM of a 220nm thick silicon disk of varying radius. The calculation is run in five distinct liquids, taken here as inviscid.

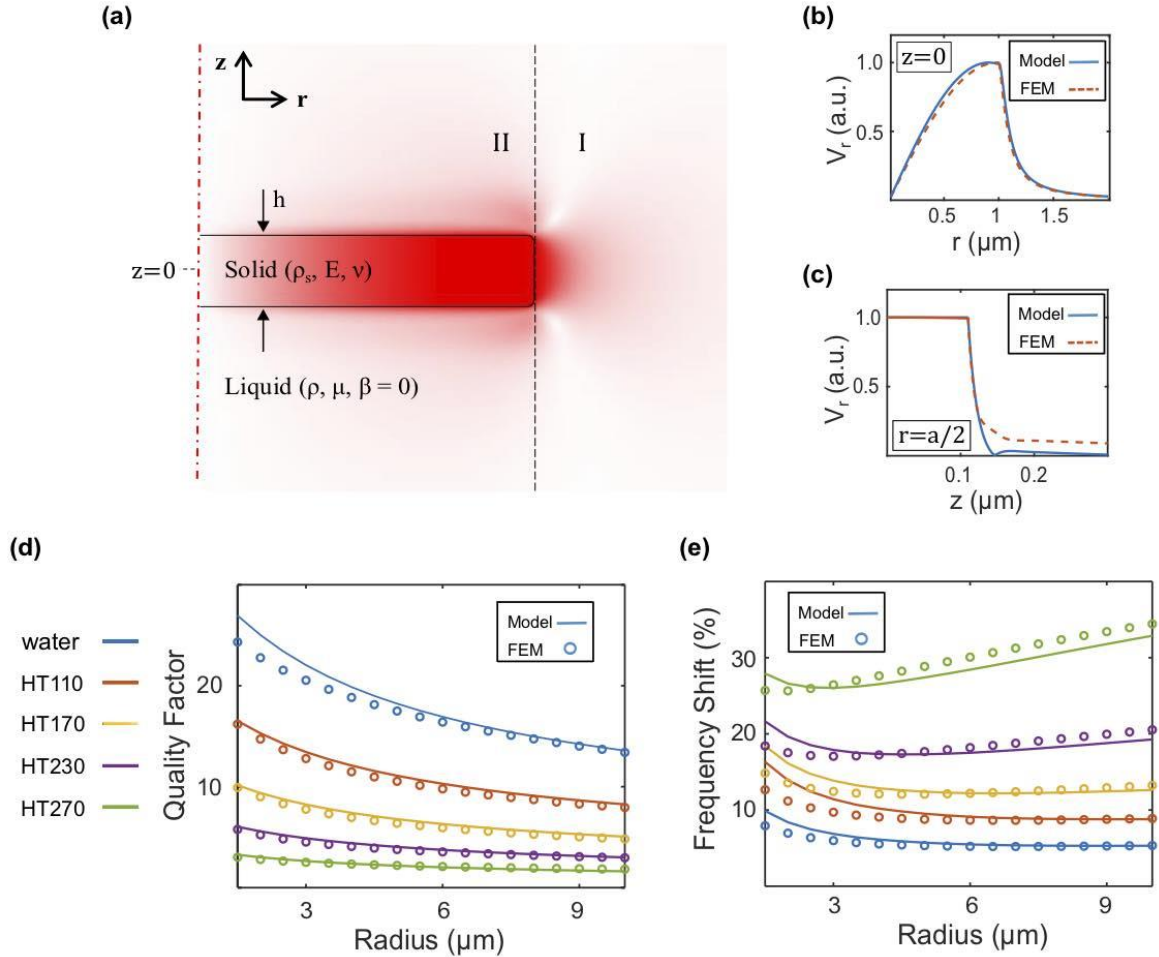

**FIG. 3.** Fluid-structure interaction for the UHF disk in the incompressible viscous fluid regime. (a) Absolute value of the radial velocity obtained by FEM. The considered situation is a silicon disk of radius a = 1μm and thickness h = 220nm, oscillating on its fundamental RBM and immersed in water, taken as incompressible and viscous. Analytically (solid line) and FEM calculated (dotted line) absolute value of the radial velocity on different intersections: (b) z = 0 and (c) r = a/2. (d) Analytically (solid line) and FEM calculated (open circles) quality factor of the disk RBM resonance versus radius, in different liquids taken as incompressible and viscous. (e) Normalized frequency shift of the RBM under the same conditions.

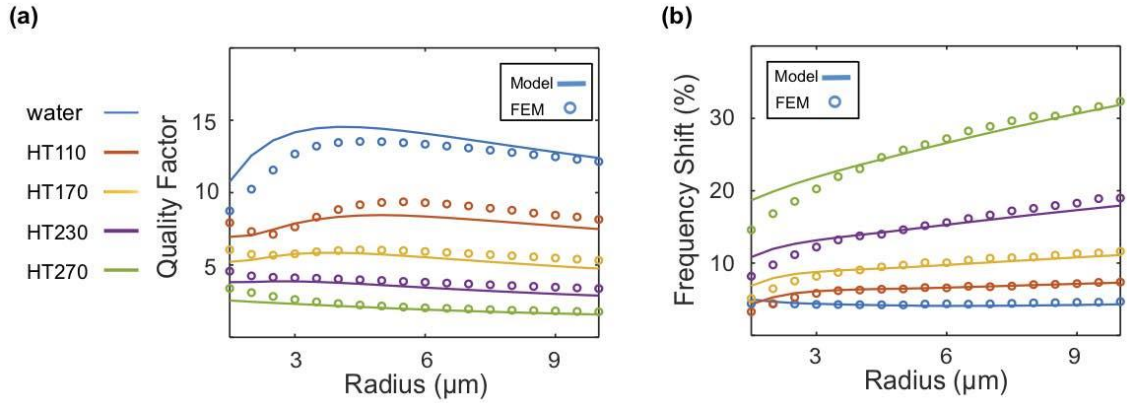

**FIG. 4.** Resonant properties of the fundamental RBM of a disk immersed in different liquids (Table1), described as compressible viscous liquids. The considered situation is a silicon disk of radius $a = 1\mu m$ and thickness $h = 220 nm$. Analytically (solid line) and FEM (round marker) calculated quality factor (a) and normalized frequency shift (b) versus radius.

# FLUID-STRUCTURE MODEL FOR DISKS VIBRATING AT ULTRA-HIGH FREQUENCY IN A COMPRESSIBLE VISCOUS FLUID: SUPPLEMENTARY MATERIAL

**I- DISK RBM IN 3D**

The general equation of motion in an isotropic, homogenous elastic solid reads:

$$\rho \frac{\partial^2 u}{\partial t^2} = (\lambda + \mu)\nabla(\nabla \cdot u) + \mu \nabla^2 u \tag{S1}$$

Where $\lambda$ and $\mu$ are the Lamé constants in 3D:

$$\lambda = \frac{\nu E}{(1-2\nu)(1+\nu)} \text{ and } \mu = \frac{E}{2(1+\nu)}.$$

We can rewrite equation S1 as:

$$\rho \frac{\partial^2 u}{\partial t^2} = (\lambda + 2\mu)\nabla(\nabla \cdot u) - \mu \nabla \times \nabla \times u \tag{S2}$$

Using the Helmholtz decomposition technique, the displacement vector is decoupled into two orthogonal components, respectively curl free and divergence free:

$$\vec{u} = -\nabla \phi + \nabla \times \vec{\psi} \tag{S3}$$

In the case of radial breathing modes, the role of $\vec{\psi}$ is negligible and the scalar potential $\phi$ suffices to depict the mechanical behavior of the elastic disk. Considering $\vec{\psi} \simeq 0$, the Navier Eq. S2 simplifies to a scalar Helmholtz equation:

$$\nabla^2 \phi + k_s^2 \phi = 0 \; ; \; k_s = \omega_s \sqrt{\frac{(1-2\nu)(1+\nu)\rho}{(1-\nu)E}} \tag{S4}$$

where $k_s$ is the wave-number and $\omega_s$ the angular frequency of vibration. This wave equation is solved by separation of variables, considering the fact that $h \ll a$. We find a scalar potential:

$$\phi = \big(A\cos(k_z z) + B\sin(k_z z)\big)J_0(k_r r) \tag{S5}$$

where $k_r^2 + k_z^2 = k_s^2$. Moreover, due to the mirror symmetry along $z$ of the radial displacement of the RBM modes, which amounts to $-\frac{\partial \phi}{\partial r}$, B should be zero. Through Eq. S3, we obtain all the components of displacement vector:

$$\mathbf{u} = A[k_r \cos(k_z z) J_1(k_r r)\mathbf{r} + k_z \sin(k_z z) J_0(k_r r)\mathbf{z}] \tag{S6}$$

This displacement field simplifies to the approximate one used in the main text when $k_z \ll k_r$.

## II- Inviscid compressible liquid (far-field approach)

For the RBM in an inviscid compressible liquid, the main source of outward acoustic wave is the sidewall of the disk, which vibrates radially with a normal velocity $v_s$. As seen in Eq. S6, this radial velocity is at first order independent of φ and z, all over the thickness h of the disk ($\cos(k_z z) \sim 1$). Therefore in the far field, the generated acoustic wave can be calculated by considering a flat ring of radius a, vibrating radially with an equivalent velocity parameter $hv_s$ (Fig. S1). Each point on that ring can be considered as generating a spherical wave $p = p_0 \frac{e^{-jkR}}{4\pi R}$, and we need to relate $p_0$ to the ring velocity. In order to circumvent mathematical singularities at the origin of a spherical wave, we replace the point source by a vibrating sphere of radius $r_0$ and radial velocity $v_r$. For this representation to lead the correct amplitude for the emitted wave in far field, one must set $v_r = \frac{hv_s}{4\pi r_0^2}$. In an inviscid fluid, the local velocity is related to the pressure through $v = -\frac{1}{j\omega\rho}\nabla p$ (for a time-varying complex velocity field $v \propto e^{j\omega t}$), hence for a spherical wave the radial component is $v_r = \frac{1}{\rho c}\left(1 + \frac{1}{jkr}\right)p$. Taking $kr_0 \ll 1$ one obtains $v_r = \frac{1}{\rho c}\left(\frac{1}{jk4\pi r_0^2}\right)p_0$, hence $p_0 = jkhv_s\rho c$. In the linear regime, the total pressure is the sum of contributions of individual sources:

$$p = p_0 \int \frac{e^{-jkR}}{4\pi R} a d\varphi'. \tag{S7}$$

where we adopted the spherical coordinates (r, φ, θ) of Fig. S1. Because we are interested in the result of a circular integral, we can set the reference of azimuthal angle φ arbitrarily. With a specific reference $\cos(\varphi) = 1$, R is expressed in the far-field as $R \approx r - a\sin(\theta)\cos(\varphi')$. The total pressure then becomes:

$$p = p_0 \frac{e^{-jkr}}{4\pi r} \int e^{jka\sin(\theta)\cos(\varphi')} a d\varphi' = p_0 a \frac{e^{-jkr}}{2r} J_0(ka\sin(\theta)) \tag{S8}$$

With the pressure known, the acoustic power radiated through the elementary surface dS of a sphere in the far field can be expressed. It is then summed over the sphere to provide the total radiated power:

$$P_{rad} = \frac{1}{2}\int Re(p^*\overrightarrow{dS} \cdot \vec{v}) = \frac{1}{2\rho c}\iint |p|^2 r^2 \sin\theta \, d\theta d\varphi \tag{S9}$$

where we used relation valid in far field $v_r = \frac{1}{\rho c} p$. At that stage, the quality factor $Q = \omega_s E_s/P_{rad}$ with $E_s$ the energy stored in the solid resonator, can be expressed in a closed form, following the procedure below:

$$u_r = Ak_s J_1(k_s r)$$

$$v_s = j\omega_s A k_s J_1(k_s a)$$

$$p_0 = j\omega_s h v_s \rho$$

Neglecting the z-displacement of the RBM, we approximate:

$$E_s \simeq \frac{1}{2}\rho_s \omega_s^2 \int |u_r^2| dV = \frac{1}{2}\rho_s \omega_s^2 \left(2\pi h k_s^2 |A|^2 \int J_1^2(k_s r) r dr\right)$$

$$= \rho_s \omega_s^2 k_s^2 \pi h |A|^2 \frac{a^2}{2}\left(J_1^2(k_s a) - J_0(k_s a)J_2(k_s a)\right)$$

$$P_{rad} = \frac{\pi}{4\rho c} a^2 |p_0|^2 \int_0^\pi J_0^2(k a \sin(\theta)) \sin(\theta) d\theta = \frac{a^2}{4c} \pi \rho h^2 \omega_s^2 |v_s|^2 \int_0^\pi J_0^2(k a \sin(\theta)) \sin(\theta) d\theta$$

$$= \frac{a^2}{4c} \pi \rho h^2 \omega_s^4 |A|^2 J_1^2(k_s a) \int_0^\pi J_0^2(k a \sin(\theta)) \sin(\theta) d\theta$$

$$Q = \frac{2c\rho_s}{\rho h \omega_s} \left(\frac{\left(1 - \frac{J_0(k_s a)J_2(k_s a)}{J_1^2(k_s a)}\right)}{\int_0^\pi J_0^2(k a \sin(\theta)) \sin(\theta) d\theta}\right) \quad (S10)$$

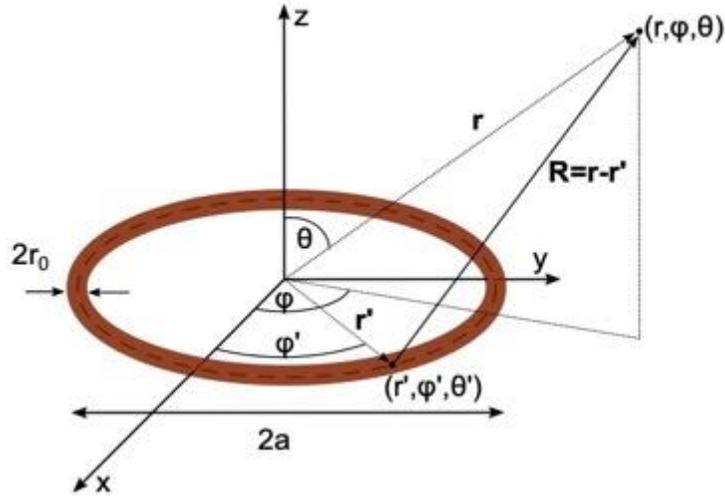

Figure S1: Flat ring vibrating radially in a liquid.

## III- Inviscid compressible liquid (near-field approach)

Using the above approach it is also possible to calculate the pressure p(r) at the surface of the loop. In order to circumvent mathematical singularities, we consider a finite thickness ($2r_0$) for the source and approximate $|\mathbf{r} - \mathbf{r}'|$ as $\sqrt{4a^2\sin^2\left(\frac{\varphi'}{2}\right) + r_0^2}$. With this approximation, we solve Eq. S7 by separating the integral in two terms:

$$p(r) = p_0 \int_0^{2\pi} \frac{e^{-jk|\mathbf{r}-\mathbf{r}'|}}{4\pi|\mathbf{r}-\mathbf{r}'|} a d\varphi' = p_0 a \int_0^{2\pi} \frac{1}{4\pi|\mathbf{r}-\mathbf{r}'|} d\varphi' + p_0 a \int_0^{2\pi} \frac{e^{-jk|\mathbf{r}-\mathbf{r}'|}-1}{4\pi|\mathbf{r}-\mathbf{r}'|} d\varphi' \qquad (S11)$$

For $r_0 \ll a$, the first term embedding the singularity is approximated by

$$p_0 a \int_0^{2\pi} \frac{1}{4\pi|\mathbf{r}-\mathbf{r}'|} d\varphi' \approx \frac{p_0}{2\pi} \ln\left(\frac{8a}{r_0}\right) \qquad (S12)$$

This term depends on $r_0$, which we will later chose to be $r_0 = \frac{h}{4}$. For the second term, with the condition $r_0 \ll a$ it is sufficient to consider $|\mathbf{r} - \mathbf{r}'| \approx 2a \cdot \sin\left(\frac{\varphi'}{2}\right)$. The integral is simplified:

$$p_0 a \int_0^{2\pi} \frac{e^{-jk|\mathbf{r}-\mathbf{r}'|}-1}{4\pi|\mathbf{r}-\mathbf{r}'|} d\varphi' = p_0 a \int_0^{2\pi} \frac{e^{-j2ka\cdot\sin\left(\frac{\varphi'}{2}\right)}-1}{4\pi\left(2a\sin\left(\frac{\varphi'}{2}\right)\right)} d\varphi' = p_0 \int_0^{\pi} \frac{e^{-j2ka\cdot\sin(\varphi')}-1}{4\pi(\sin(\varphi'))} d\varphi' \qquad (S13)$$

We further express this integral in terms of Bessel and Struve functions:

$$p_0 \int_0^{\pi} \frac{e^{-j2ka\cdot\sin(\varphi')}-1}{4\pi(\sin(\varphi'))} d\varphi' = \frac{-jp_0}{4\pi} \int_0^{2ka} dx \int_0^{\pi} e^{-jx\cdot\sin(\varphi')} d\varphi' = \frac{-jp_0}{4} \int_0^{2ka} (J_0(x) - jH_0(x)) dx \qquad (S14)$$

where $J_0(x)$, $H_0(x)$ are Bessel and Struve functions:

$$J_0(x) = \frac{1}{\pi} \int_0^{\pi} \cos(x\sin(\varphi)) d\varphi \qquad (S15)$$

$$H_0(x) = \frac{1}{\pi} \int_0^{\pi} \sin(x\sin(\varphi)) d\varphi \qquad (S16)$$

Inserting (S12) and (S10) into (S9) leads to:

$$p(r) = \omega \rho h v_s \left(\frac{j}{2\pi} \ln\left(\frac{8a}{r_0}\right) + \frac{1}{4} \int_0^{2ka} (J_0(x) - jH_0(x)) dx\right). \qquad (S17)$$

Finally, with p and $v_s$ at our disposal, we can calculate the modified mass and the quality factor, following Eqs (4) and (5) of the main text:

$$m_r = \left(1 + \frac{\text{Im}\left(\int_0^{2\pi} pv_s^* a 2r_0 d\varphi\right)}{2\omega_s E_s}\right) = 1 + \frac{2\rho h\left(\frac{1}{2\pi}\ln\left(\frac{8a}{r_0}\right) - \frac{1}{4}\int_0^{2ka} H_0(x) dx\right)}{a\rho_s\left(1 - \frac{J_0(k_s a)J_2(k_s a)}{J_1^2(k_s a)}\right)} \qquad (S18)$$

$$Q = \frac{2\omega_s E_s}{\text{Re}\left(\int_0^{2\pi} pv_s^* a 2r_0 d\varphi\right)} = \left(\frac{2a\rho_s\left(1 - \frac{J_0(k_s a)J_2(k_s a)}{J_1^2(k_s a)}\right)}{\rho h\left(\int_0^{2ka} J_0(x) dx\right)}\right) \qquad (S19)$$

The expression obtained in (Eq. S19) through a near-field calculation seems to differ from that obtained in the far field (Eq. S10). In figure S2, we check numerically that these two expressions actually match for our considered disk resonators and for liquids of interest.

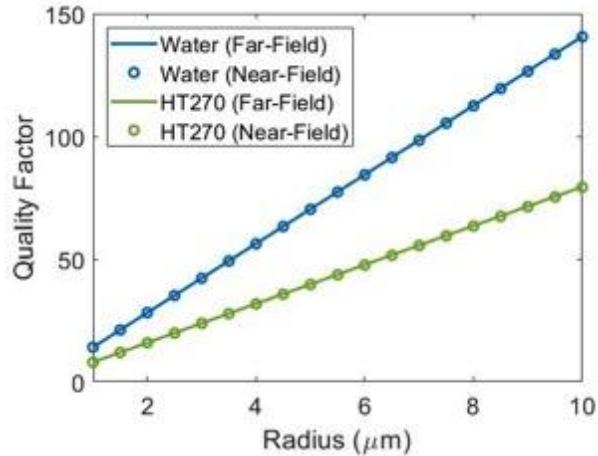

Figure S2: Comparison of closed-form expressions for Q in an inviscid liquid. Expressions (S10) and (S19) are compared for a varying radius of the disk, and for two distinct liquids: water and the perfluorinated liquid HT270. There is no noticeable deviation between both expressions.

## IV- Inviscid compressible liquid: comparison of approaches

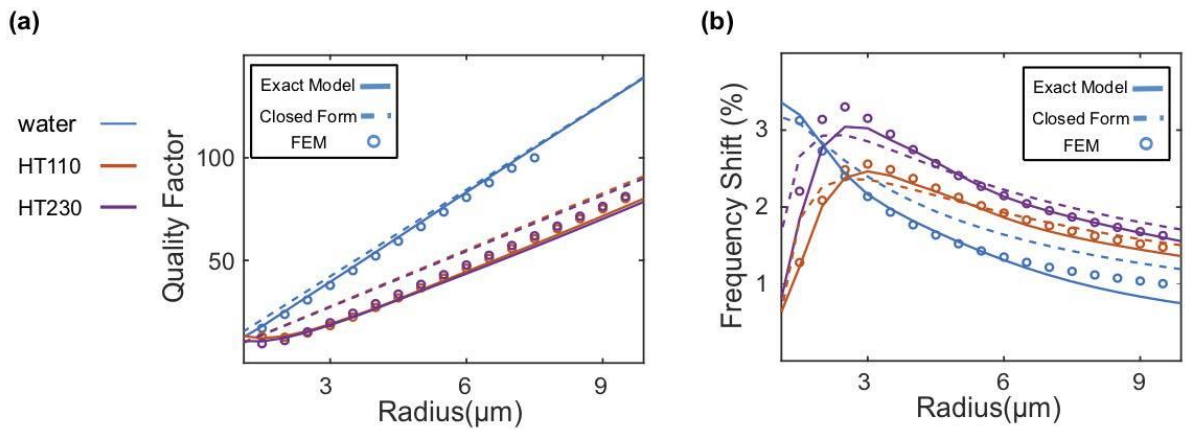

Figure S3: The results of our three distinct modeling approaches are compared for a disk of varying radius immersed in three distinct liquids: water and two perfluorinated liquids HT110 and HT 230. The first analytical approach introduced in the main text is dubbed "exact model" (solid line), and contrasts with the approximated closed-form results obtained above (dashed line, Eqs. S19 and S18). The FEM results are shown as open symbols, and come very close to the exact model results. The closed-form expressions provide very efficient description of the trends, and a quite satisfactory quantitative agreement as well, with deviation that remains below 30% even in the worst cases.

## V- Incompressible Viscous Liquid

In the incompressible viscous case and for our particular geometry, we can write down approximated boundary conditions for the scalar and vector potential at solid-liquid interfaces of the problem, as shown below. With such approximation, one can solve for the two potentials separately (Eqs. 12 and 13 of main text) and obtain the complete fluidic response using the superposition principle.

We separate variables in order to find a solution for $\psi$, the azimuthal component of the vector potential: $\psi(r,z) = R(r)Z(z)$. The differential equations read:

$$\frac{d^2 Z(z)}{dz^2} + \lambda^2 Z(z) = 0 \tag{S20}$$

$$\frac{d^2 R(r)}{dr^2} + \frac{1}{r}\frac{dR(r)}{dr} + \left(\sigma^2 - \lambda^2 - \frac{1}{r^2}\right) R(r) = 0 \tag{S21}$$

with $\lambda^2$ a constant, positive or negative. Equation S21 adopts first-order Bessel functions of argument $k_r r$ as solutions, with $k_r^2 = \sigma^2 - \lambda^2$. Since $\sigma^2$ is imaginary, the solutions are attenuated waves. As a result, the radial velocity decays far from the disk, as shown by FEM calculations in Fig. S4.

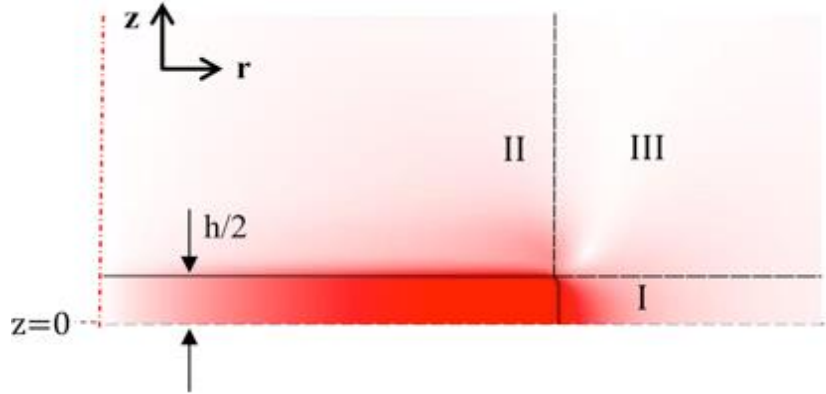

Figure S4: Typical profile of the norm of the radial velocity. The considered fluid-structure configuration is a thin disk vibrating on its fundamental RBM in an incompressible viscous fluid.

In consequence of this rapid decay, distinct fluid domains can be advantageously considered. Domains I, II and III are defined graphically in Fig. S4. In domain III the fluid velocity comes close to zero, and there is negligible interaction between domains I and II, which are physically separated by domain III. As a result, the hydrodynamic problem can be separated in two independent problems in domains I and II. In domain I, the continuity of the velocity at the solid-fluid interface (r=a, $|z|$<h/2) must be satisfied, while free boundary conditions can be considered at the interface with domain III (r>a,$|z|$=h/2). In domain II, a continuity of the velocity at the solid-fluid interface ($|z|$=h/2, r<a) must be satisfied, while free boundary conditions can be considered at the interface with domain III (r=a, |z|>h/2). In this description, domains I and II have a finite extension in the transverse direction (orthogonal to the propagation of the waves), hence a finite series of basis functions can be employed to solve Eqs. 12 and 13 of the main text (instead of an integral of functions in the case of an infinite interval). In domain I, the continuity of velocity at the solid-fluid interface (r=a) simplifies to a continuity of radial (normal) velocity, as the vertical (z) velocity at the disk sidewall is negligibly smaller (see Fig. S5). Similarly in domain II, the continuity of the velocity at the solid-fluid interface (z=h/2, r<a) simplifies to a continuity of radial (tangential) velocity, as the vertical (z) velocity at the disk upper surface is negligible (see Fig. S5). Under these simplifications, the radial velocity at the disk-fluid interfaces determines the series of basis functions that need to be considered. In the particular case of the 1st RBM mode of the disk, the radial velocity at the disk sidewall (r=a) is almost independent of z over the disk thickness. More precisely, it is well approximated by a single $\cos(k_z z)$ function with $k_z h \ll 1$ (later below, we will take $k_z=0$). In that limit a single term $k_{zI} = k_z$ of the z series expansion is required to describe $\psi$ in domain I:

$$\psi_I = A_I H_1^{(2)}(k_{r_I} r) \sin(k_{z_I} z); \quad k_{r_I}^2 + k_{z_I}^2 = \sigma^2 \tag{S22}$$

with $k_{z_I}$ real and small ($k_{zI} h \ll 1$), and $k_{r_I}$ complex with a negative imaginary part.

In domain II also, for the particular case of the 1st RBM mode of the disk, the radial velocity at the disk interface z=h/2 is set by $J_1(k_s r)$ and a single term of the radial series expansion ($k_{r_{II}} = k_s$) is hence required to obtain the potential $\psi$:

$$\psi_{II} = A_{II} J_1(k_{r_{II}} r) e^{-k_{zII}(|z|-\frac{h}{2})}; \quad k_{r_{II}}^2 - k_{z_{II}}^2 = \sigma^2 \tag{S23}$$

with $k_{r_{II}} = k_S$ real-valued and $k_{zII}$ complex with a positive real part. Knowing that $v_{r\phi}$ (the radial velocity associated to the $\phi$ potential) is negligible at the top and bottom surface of the disk (Fig. S5), the continuity of radial velocity at the fluid-structure interfaces can be expressed as:

$$\left(-\frac{\partial \psi_{II}}{\partial z}\right)_{r<a, |z|=h/2} = (j\omega_s u_r)_{r<a, |z|=h/2} \tag{S24}$$

$$\left(\frac{1}{r}\frac{\partial (r\psi_I)}{\partial r}\right)_{r=a,|z|<h/2} + \left(\frac{\partial \phi_I}{\partial z}\right)_{r=a,|z|<h/2} = (j\omega_s u_z)_{r=a,|z|<h/2} = 0 \tag{S25}$$

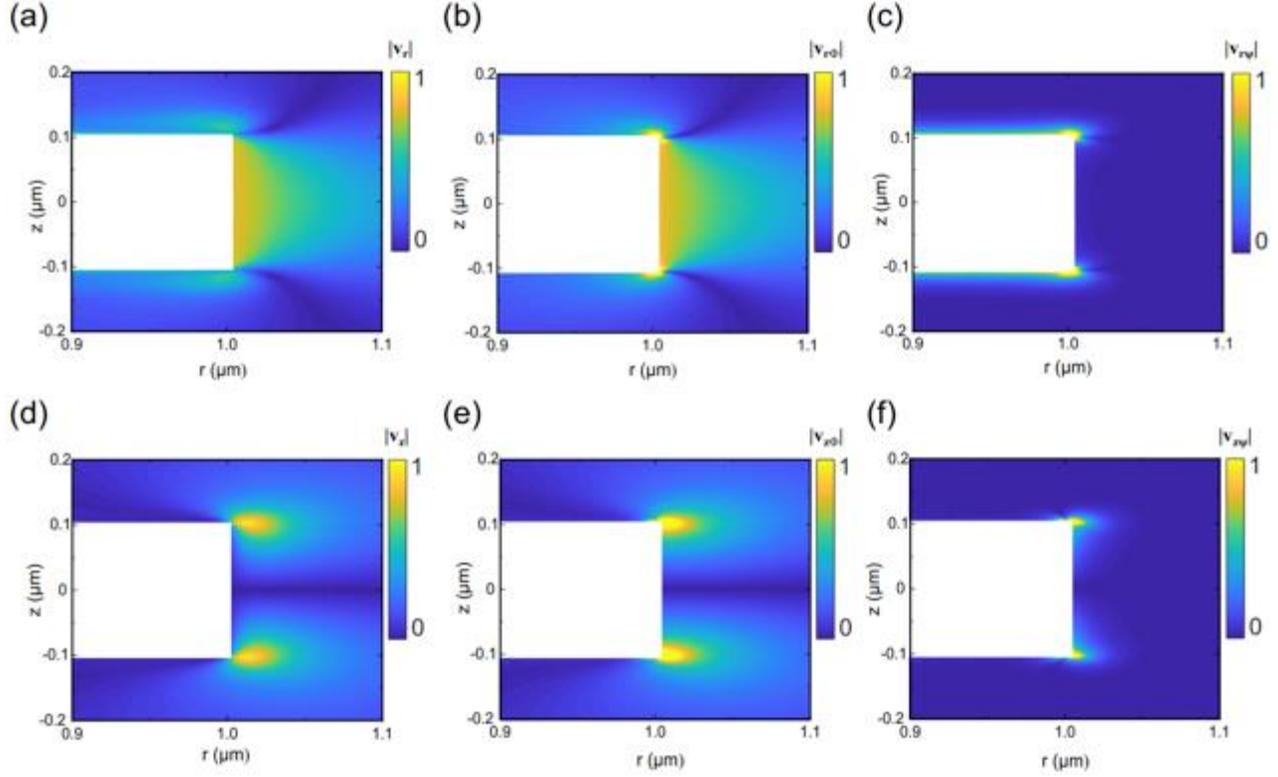

Fig. S5. Helmholtz decomposition of the fluid velocity field in the incompressible viscous case, obtained by FEM. The fluid velocity around a silicon disk of radius a=1µm and thickness h=220nm, vibrating on its fundamental RBM, is decomposed into its curl-free ($v_\phi$) and divergence-free ($v_\Psi$) components. (a-c) Radial velocity and its two Helmholtz decomposition components. (d-f) Vertical (z) velocity and its two Helmholtz decomposition components.

Notice that $k_s^2 \ll |\sigma^2|$ for UHF mechanical resonators made out of a typical elastic semiconductor material, and for liquid viscosities of interest below the Pa s. With Eqs. S23, S24 and S25, this allows obtaining the (radial) velocity on the disk top surface (interface with domain II), together with the stress component required to compute Eq. 2 of the main text:

$$v_r = -\frac{\partial \psi_{II}}{\partial z} = j\omega_s A k_s J_1(k_s r) e^{-j\sigma(|z|-\frac{h}{2})} \tag{S26}$$

$$\sigma_{rz} = \mu \frac{\partial v_r}{\partial z} = -j\sigma\mu v_r \tag{S27}$$

With these two quantities, which are the same on the top and bottom surface of the disk, we can compute the contribution of these interfaces in the evolution of the disk vibration:

$$\frac{2(\int v_r^* \sigma_{rz} r dr d\varphi)}{2\omega_s E_s} = \frac{-2j\mu\sigma}{\omega_s \rho_s h} \tag{S28}$$

where we neglected the z-component of motion in the mechanical energy of the vibrating disk.

In order to compute the sidewall contribution, we need to carry a similar analysis at the interface with domain I. In domain I, at the interface since tangential velocity $v_z$ is negligible:

$$\int v_j^* \sigma_{ij} n_i d\Sigma = \int v_r^* \sigma_{rr} a d\varphi dz \tag{S29}$$

With $\sigma_{rr} = -p + 2\mu \frac{dv_r}{dr}$. As can be seen in Fig. S5, on the disk sidewall $v_r \approx v_{r\phi}$. The pressure p at the sidewall interface in the incompressible case (k=0) can be calculated using Eq. S17. In order to estimate $\frac{dv_r}{dr}$ we neglect near the corners the field singularities visible in Fig. S5 magnified in Fig. S6. Relying on Fig. S6 we approximate the profile of $\mathbf{v}_\phi$ on the sidewall surface by:

$$v_{r\phi} = v_s \tag{S30}$$

$$v_{z\phi} = v_s \sin(k_z z) \tag{S31}$$

with $v_s = j\omega_s A k_s J_1(k_s a)$ and $k_z = \frac{\pi}{h} \ll |\sigma|$.

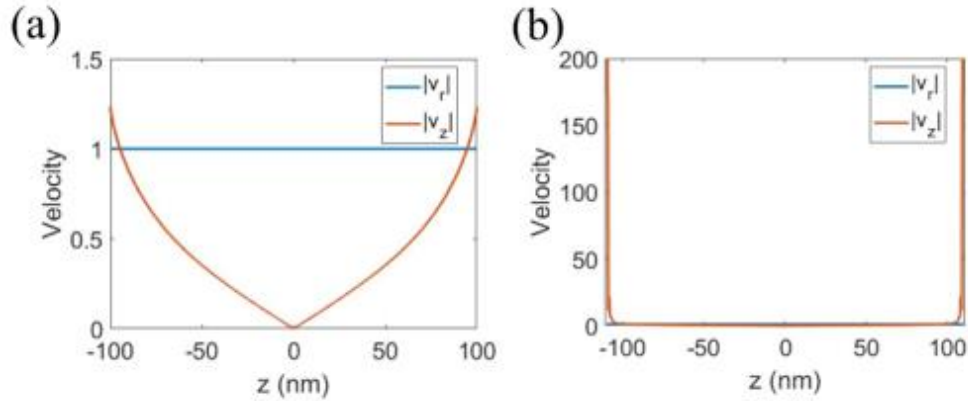

Fig. S6. The curl-free component of the velocity (vΦ) at the surface of the disk sidewall (r=a). Silicon disk of radius a=2 μm and thickness h=220 nm immersed in incompressible water. (a) |z|<100nm, (b) |z|<110nm.

Eq. S31 means that a single mode is sufficient for our description, hence according to equations 14 and 15 of main text, one can approximate:

$$v_{r\phi} = -A \cos(k_{zI}z)\, k_r H_1^{(2)}(k_r r)$$

Since $k_{zI}z \le 1$, $v_{r\phi} = -A\, k_r H_1^{(2)}(k_{rI}r)$ and A can be calculated at the interface where $v_{r\phi} = v_s$. In domain I, we finally obtain:

$$v_r \approx v_{r\phi} = v_s \frac{H_1^{(2)}(k_{rI}r)}{H_1^{(2)}(k_{rI}a)} \quad (S32)$$

Substituting $v_{r\phi}$ in Eq. S29, we obtain the contribution of the sidewall (r=a, |z|<h/2) to the quantity of interest:

$$\frac{\left(\int v_r^* \sigma_{rr} a d\varphi dz\right)}{2\omega_s E_s} = -\frac{2\mu \frac{k_I (H_2^{(2)}(k_{rI}a) - H_0^{(2)}(k_{rI}a))}{\omega_s H_1^{(2)}(k_{rI}a)} + \frac{j\rho h}{\pi} \ln\left(\frac{8a}{r_0}\right)}{a\rho_s \left(1 - \frac{J_0(k_s a) J_2(k_s a)}{J_1^2(k_s a)}\right)} \quad (S33)$$

The total energy evolution due to fluid-structure interactions on all interfaces is finally obtained:

$$\frac{\left(\int \tilde{v}_j \tilde{\sigma}_{ij} n_i d\Sigma\right)}{2\omega_s E_s} = \frac{-2j\mu\sigma}{\omega_s \rho_s h} - \frac{2\mu \frac{k_{rI}(H_2^{(2)}(k_{rI}a) - H_0^{(2)}(k_{rI}a))}{\omega_s H_1^{(2)}(k_{rI}a)} + \frac{j\rho h}{\pi}\ln\left(\frac{8a}{r_0}\right)}{a\rho_s\left(1 - \frac{J_0(k_s a)J_2(k_s a)}{J_1^2(k_s a)}\right)} \quad (S34)$$

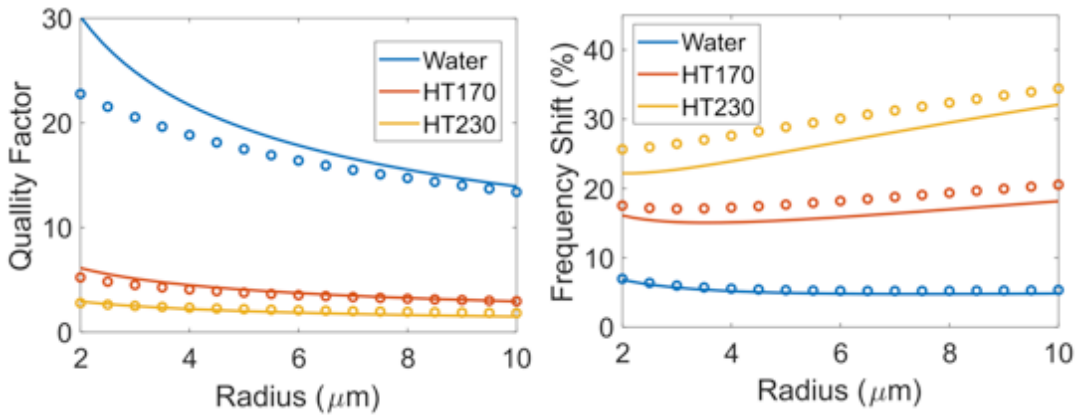

Fig. S7. Calculated quality factor and frequency shift using closed-form approximation from Eq. S34 (solid line) and FEM (open circle). Silicon disk of thickness h=220 nm, immersed in a liquid.

As can be seen in Fig. S7, the results of the closed-form approximation Eq. S34 reproduce well the evolution of both quality factor and frequency shift. This said, they slightly deviate from FEM simulations at small radius. The reason for this is the singularity near the corners. To achieve better agreement, we can include the effect of the corner by empirically modifying $k_r$ in Eq. S34.

## VI- Compressible viscous liquid

Following the path established in previous section, we can now treat the case of an arbitrary compressible viscous liquid by employing the global expression of pressure derived in Eq. S17:

$$\frac{\left(\int \tilde{v}_j \tilde{\sigma}_{ij} n_i d\Sigma\right)}{2\omega_s E_s} = -\frac{\sqrt{2}(1+j)\sqrt{\frac{\mu\rho}{\omega_s}}}{h\rho_s} - \frac{2\mu \frac{k_{rI}(H_2^{(2)}(k_{rI}a)-H_0^{(2)}(k_{rI}a))}{\omega_s H_1^{(2)}(k_{rI}a)} + \rho h\left(\frac{j}{\pi}\ln\left(\frac{32a}{h}\right)+\frac{1}{2}\int_0^{2ka}(J_0(x)-jH_0(x))dx\right)}{a\rho_s\left(1-\frac{J_0(k_s a)J_2(k_s a)}{J_1^2(k_s a)}\right)} \quad (S35)$$

where we have expressed the first term of Eq. S34 in a different manner. In the main text, because $k_{zI} \ll |\sigma|$, we consider $k_{rI} = \sigma$ in order to write down Eqs. 24 and 25. In Fig. S8, the results of Eq. S35 are compared with FEM calculations, for three distinct liquids. There again the agreement is very good, with residual deviations at small radius for a low viscosity liquid such as water.

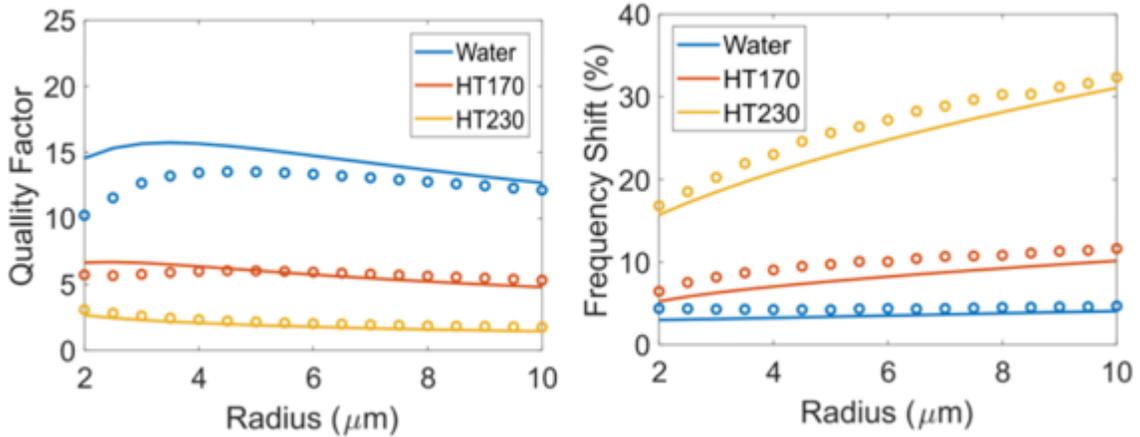

Fig. S8. Quality factor and frequency shift of the first order RBM, calculated using closed-form approximation Eq. S35 (solid line) and FEM (open circle).

Since $k_{r_I}a \gg 1$, the expression $\frac{(H_2^{(2)}(k_{r_I}a) - H_0^{(2)}(k_{r_I}a))}{H_1^{(2)}(k_{r_I}a)}$ simplifies to 2j. Using this simplification, we obtain convenient expressions for $m_r$ and $\gamma$:

$$m_r = 1 + \frac{\sqrt{\frac{2\mu\rho}{\omega_s}}}{\rho_s h} + \frac{2\sqrt{\frac{2\mu\rho}{\omega_s}} + \rho h \left(\frac{1}{\pi} ln\left(\frac{32a}{h}\right) - \frac{1}{2}\int_0^{2ka} H_0(x)dx\right)}{a\rho_s \left(1 - \frac{J_0(k_s a) J_2(k_s a)}{J_1^2(k_s a)}\right)}$$

$$\gamma = \frac{\sqrt{2\mu\rho\omega_s}}{\rho_s h} + \frac{\sqrt{2\mu\rho\omega_s} + \frac{\omega_s \rho h}{2}\int_0^{2ka} J_0(x)dx}{a\rho_s \left(1 - \frac{J_0(k_s a) J_2(k_s a)}{J_1^2(k_s a)}\right)}$$

For completeness, we show in Fig. S9 the Helmholtz decomposition of the velocity field around the disk for a compressible viscous liquid.

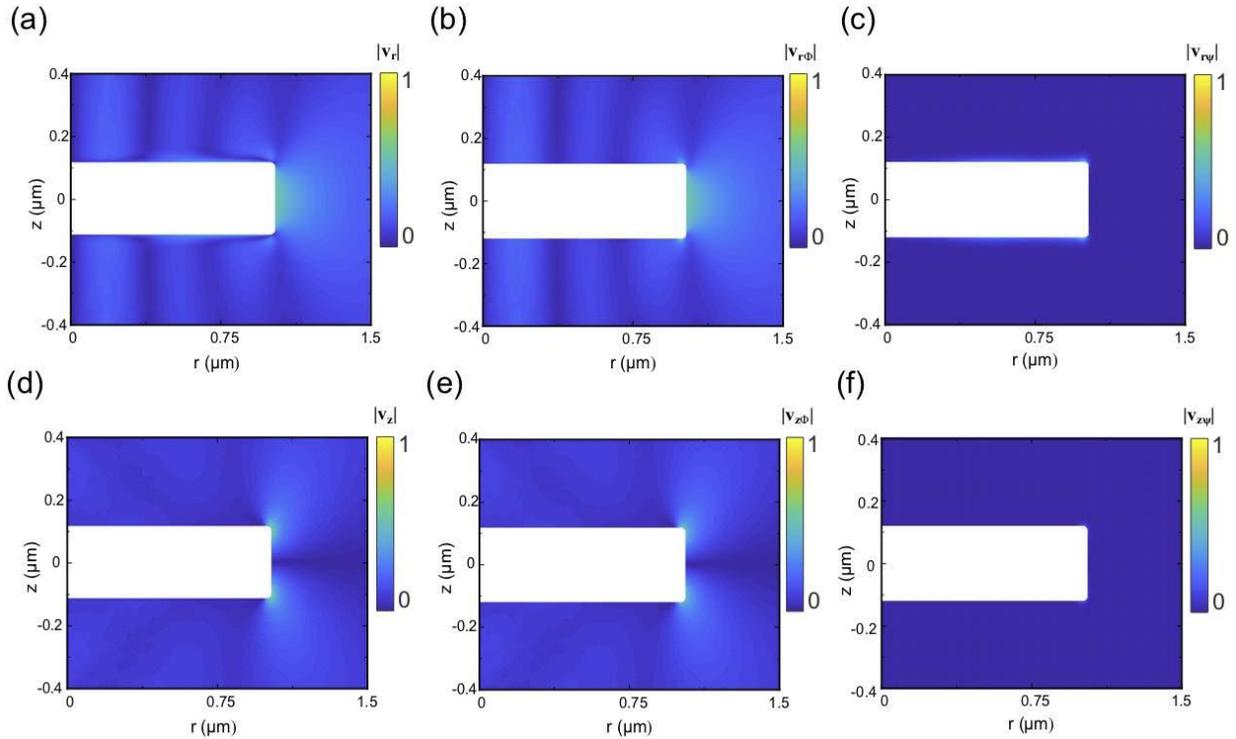

Fig. S9. Helmholtz decomposition of the velocity field in the compressible viscous case. The velocity around a silicon disk of radius a=1 μm and thickness h=220 nm is decomposed into curl-free (Φ) and divergence-free (Ψ) components. (a-c) Radial velocity and its components. (d-f) Vertical (z) velocity and its components.

## VII- Extension to the case of finite bulk viscosity

Following the previous sections path, we obtain the total energy evolution when $\mu_B \neq 0$:

$$j\omega_s \rho \mathbf{v} = -\frac{1}{j\omega_s \rho} \nabla p \left(1 + \left(\frac{4}{3}\mu + \mu_B\right) j\omega_s \beta\right)$$

$$p_0 = \frac{j\omega_s \rho h v_s}{1 + j\omega_s \beta(\mu_B + \frac{4}{3}\mu)}$$

$$\frac{\left(\int \tilde{v}_j \tilde{\sigma}_{ij} n_i d\Sigma\right)}{2\omega_s E_s} = -\frac{2j\mu\sigma}{\omega_s \rho_s h} - \frac{2\mu \frac{k_{rI}(H_2^{(2)}(k_{rI}a) - H_0^{(2)}(k_{rI}a))}{\omega_s H_1^{(2)}(k_{rI}a)} + \frac{\rho h\left(\frac{j}{\pi}\ln\left(\frac{32a}{h}\right) + \frac{1}{2}\int_0^{2ka}(J_0(x) - jH_0(x))dx\right)}{1 + j\omega_s \beta(\mu_B + \frac{4}{3}\mu)}}{a\rho_s \left(1 - \frac{J_0(k_s a) J_2(k_s a)}{J_1^2(k_s a)}\right)} \quad (35)$$

With $k_z = \frac{\pi}{h}$, $k = \omega_s \sqrt{\frac{\beta \rho}{1 + j\omega_s \beta\left(\frac{4}{3}\mu + \mu_B\right)}}$, and $k_{rI}^2 + k_z^2 = \sigma^2$.

## VIII- FEM Simulations

We simulated a vibrating silicon disk resonator in a compressible viscous fluid using the Acoustics module of the Comsol software. Taking advantage of the axial symmetry of the disk, we used 2D axisymmetric models, imposing no-slip boundary conditions at the solid-liquid interfaces. Perfectly matched layers were used to prevent reflections at the outer boundaries of the computation region. The spherical computational region was chosen to be large enough to avoid undesirable interactions between the disk and the outer boundary. Convergence was checked during simulations. As a rule of thumb, the mesh size was required to be smaller than $\lambda/10$.